# Conscious Perception: Time for an Update?

Moti Salti[1,2], Asaf Harel[1,2], Sébastien Marti[3]

1. Brain Imaging Research Center, Ben-Gurion University and Soroka University Hospital, Beer-Sheva, Israel
2. Zlotowski Center for Neuroscience, Ben-Gurion University of the Negev, Beer-Sheva, Israel
3. Cognitive Neuroimaging Unit, CEA DSV/I2BM, INSERM, Université Paris-Saclay, NeuroSpin Center, 91191 Gif/Yvette, France

Corresponding Author:

Moti Salti

Email: motisalti@gmail.com

# Abstract (150 words)

Understanding the neural mechanisms underlying subjective representation has become a central endeavor in cognitive-neuroscience. In theories of conscious perception, stimulus gaining conscious access is usually considered as a discrete neuronal event to be characterized in time or space, sometimes referred to as a conscious 'episode'. Surprisingly, the alternative hypothesis according to which conscious perception is a dynamic process has been rarely considered. Here, we discuss this hypothesis and envisage its implication. We show how it can reconcile inconsistent empirical findings on the timing of the neural correlates of consciousness (NCCs), and make testable predictions. According to this hypothesis, a stimulus is consciously perceived for as long as it is recoded to fit an ongoing stream composed of all other perceived stimuli. We suggest that this 'updating' process is governed by at least three factors (1) context, (2) stimulus saliency, and (3) observer's goal. Finally, this framework forces us to reconsider the typical distinction between conscious and unconscious information processing.

# 1. Introduction

The main thrust of the scientific study of conscious perception is to explain the cognitive processes that make a piece of external or internal information available for subjective experience. In a seminal paper Crick and Koch suggested that characterizing the neural activity that underlies conscious perception would advance our understanding of subjective experience (Crick & Koch, 1990). In a typical NCC experiment, while the neural activity is recorded, degraded stimuli (note that most of NCC studies were done in the visual domain) are presented. Participants are asked whether they perceived the target or not. NCCs are extracted with the contrast of the brain activity associated with each of these conditions. The study of NCC depends on delineating the boundaries between conscious and unconscious perception (for review Dehaene & Changeux, 2011). However, in almost three decades this approach has been practiced, there is no consensus regarding the boundaries between conscious and unconscious perception and consequently regarding the neural activity underlying conscious and unconscious perception.

Our goal here is not to propose yet another model of consciousness but simply to make a point: Should we re-think what we know about conscious perception? Finding "when" consciousness arises already makes an assumption: Conscious perception corresponds to a specific brain event, a period of stability in the neuronal activity reflecting subjective experience. Here, we point out that the alternative hypothesis should be considered as well: Conscious perception might correspond to a dynamic process involving a series of brain processing stages. In the following, we consider the implications of this hypothesis, confront it to extant data, and draw testable predictions.

# 2. 'Early' versus 'Late' NCCs: An unsolvable debate?

The main endeavor of the NCC study is to isolate the minimal spatial and temporal brain activity that gives rise to conscious perception. However, to date there is no consensus regarding the activity that gives rise to consciousness. An Activation Likelihood Estimation (ALE) meta-analysis performed recently Bisenius and colleagues (Bisenius, Trapp, Neumann, & Schroeter, 2015) yielded diffused activity that included clusters of activation in inferior and middle occipital gyrus; fusiform gyrus; inferior temporal gyrus; caudate nucleus; insula; inferior, middle, and superior frontal gyri; precuneus; as well as in inferior and superior parietal lobules. In the temporal domain, as the explored space is more restricted, the P3b component (~350 ms after stimulus onset) and a negative early component, usually referred to as Visual Awareness Negativity (VAN, ~200 ms after stimulus onset) stand-out as potential NCCs (Koivisto & Revonsuo, 2003; Koivisto, Salminen-Vaparanta, Grassini, Revonsuo, & Foxe, 2016). According to the taxonomy put forward by Aru (Aru, Bachmann, Singer, & Melloni, 2012) and colleagues , the genuine NCC should be distinguished from (i) events of that precede and support consciousness "NCC-pr", and (ii) events that follow consciousness and are the consequence of it "NCC-co". The empirical results in the temporal and spatial domain are utilized to support two main hypotheses. According to the early entry hypothesis, (visual) conscious perception relies on early activity in striate and extrastriate activity and is reflected by the VAN ERP component. The P3b would only reflect response-related processes that follow conscious perception. Using the inattentional blindness paradigm, Pitts and colleagues showed that the conscious perception of an expected stimulus irrelevant to the task is associated with a larger amplitude of the VAN rather than the P3b. Only once the target stimulus became task-relevant does the P3b reflect conscious perception. This result was consistent with earlier findings in studies examining the effects of expectation on conscious perception. These studies revealed that when a target stimulus is seen and unexpected, a large P3b component is observed. By contrast, when the same stimulus is perceived but expected, a difference is observed at earlier latency (e.g. on the N2 (Mika Koivisto & Revonsuo, 2008) or the P2 components (Melloni, Schwiedrzik, Müller, Rodriguez, & Singer, 2011)) but no effect is observed in the P3b time window.

In contrary, the late access hypothesis asserts that conscious perception relies on higher activity that includes parietal and prefrontal cortices and is reflected by the P3b ERP component. Many studies in humans and monkeys showed that visual processing up to ~250 ms, and thus including the VAN, is preserved even when subjects deny having seen a target stimulus (Del Cul, Baillet, & Dehaene, 2007; Kovacs, Vogels, & Orban, 1995; Lamme, Zipser, & Spekreijse, 2002; Marti, Sigman, & Dehaene, 2012; Rolls, Tovee, & Panzeri, 1999; Sergent, Baillet, & Dehaene, 2005; Vogel, Luck, & Shapiro, 1998). The VAN would therefore manifest processes that precede and support consciousness, like attention for instance (Busch, Frund, & Herrmann, 2010). This dispute on the exact timing of NCCs thus seems unsolvable and in fact, in many experiments 'early' and 'late' correlates of consciousness co-exist ( Railo, Koivisto, & Revonsuo, 2011 but see Koivisto, Revonsuo, & Lehtonen, 2006;).

## 3. A broader look of the NCC in the general ERP literature

One way to untie what seems to be a Gordian knot is to examine these components role in the general ERP literature and dynamics and their relations between these results. The P3b is one of the most studied ERP component, it can be easily elicited with an oddball paradigm in which subjects are presented with low-probability target items that are mixed with high-probability non-target (or "standard") items, both perfectly visible. The P3b is clearly observed after the presentation of a target but is barely measurable after standard items. Most prominent and generalized theory attributes the P3b to Context Updating (Donchin & Coles, 1988) which relates to the changes of perception of one's surroundings, or the revision of the subject's mental representation of the situation around him rather than acting as a call to action for a certain response, be it motor, verbal, or otherwise. The context-updating theory asserts that after initial sensory processing, a comparison mechanism checks for changes from previous events in working memory. If there is no stimulus attribute change, the current mental schema remains, evoking only the sensory evoked potentials (N1, P2, N2). Importantly, only if a novel stimulus (either in content or in characteristics) is detected, and enough attentional resources are present, an "updating" of the stimulus representation occurs, eliciting the P3b potential (Polich, 2007).

A recent study conducted by Salti and colleagues makes a bridge between the interpretations of the P3b in the general ERP literature and what is observed in NCC studies. The authors contrasted the time-resolved performance of a multivariate classifier for "Seen" and "Unseen" trials in extracting the content of perception from M/EEG (Salti et al., 2015). As expected classification was superior in 'Seen' trials than in 'Unseen' trials 250-800 ms post stimulus presentation. However, this improved classification did not stem from a recruitment of additional brain regions in the encoding of consciously perceived stimulus, nor was this information coded for a longer duration. Instead, the encoding of "Seen" and "Unseen" stimuli differed in their moment-to-moment dynamics. An analysis aiming at testing the dynamics and transient stability of internal codes, used the temporal generalization method (King & Dehaene, 2014). With this method, the efficiency of a classifier trained at a certain time point t can generalize to other time points, t'. If the representation is stable, the classifier should remain efficient even if applied at a different latency. This analysis showed that the encoding of "Unseen" stimuli was much more stable and exhibited a slow decay of ~350ms. By contrast, the encoding of "Seen" trials consisted in a series of patterns of activity rapidly changing after ~160ms. We suggest these results reflect a moment-to-moment encoding mechanism. Accordingly, for reported 'Seen' stimuli information was systematically recoded via a chain of short-lived processes. For stimuli reported as 'Unseen', however, this dynamic recoding was interrupted, resulting in a slow decay of the current stimuli coding.

## Conscious perception: A moment-to-moment updating process

Adopting the perspective of a moment-to-moment (MTM) updating model, conscious perception of a stimulus is not a singular event but instead a continuous process which encompass past perceptual events and adjusts the system for future, predicted ones. In this context, the information that is being processed is part of the subjective experience. To elaborate, subjective perception, from this perspective, entails an ongoing reconstruction of the outside surroundings to an internal representation. We point out two putative parameters that guide this update as reflected in the extant data. The first parameter is the observer's goals. The MTM model interprets the lack of P3b in the recent inattentional blindness studies by Pitts and colleagues as a failure to update the target stimulus or a very short update as it does not serve a predefined top-down purpose. Changes to the existing context rely on internal goals and stimuli saliency. The goals modulate the representations of stimuli, prioritizing them and determining their availability for report, action or other cognitive processes. Some of these contents stay activated in the perceptual system until they are exhausted while others fade away. In 'Exhausted' we refer to a situation in which information of these contents served the goals for which they were prioritized. For the contents that endure, encoding evolves and changes to keep them coherent with the visual scene. The duration of this continuous coding processes, or the life-span of a stimulus in subjective experience, is not constant and depends not only on the subject's goals (e.g. subjective report, speeded response) but also on bottom-up factors (e.g. saliency). This framework predicts that the temporal modulation of the context by the observer's goals and stimulus saliency should affect the observer's change in perceived contents.

Let's consider an example: a subject is presented an image and is instructed to produce a motor response. The task is accomplished through a series of cognitive operations, from basic visual processing to motor planning. For each of these steps, different populations of neurons are involved. Every time a new population of neurons is recruited, the internal information is updated and the subjective experience changes. This implies that subjective experience evolves in time and is directly related to the successive stages of neuronal processing. The temporal evolution of subjective experience might not be linear: If multiple brain areas are simultaneously recruited – as in the P3b time window – then the complexity of the representation should drastically increase, and the subjective experience should change accordingly. Within a certain brain area or at the level of a neuronal population, the pattern of activity can remain stable (Schurger, Pereira, Treisman, & Cohen, 2010). However, at the brain level, the activity is continuously evolving and the information shared across areas evolves as well.

In many aspects this framing of subjective perception corresponds to the ideas that were elaborated in the more general framework of predictive coding. This framework which suggests that the brain builds a model of the environment based on past sensory information and uses it to predict upcoming events. If the predictive model is accurate, the difference between the new sensory input and the internal model, the 'prediction error', is minimal. On the other hand, if the difference between the two is important, then the model is inaccurate and needs to be revised. Once the prediction error is minimized, then the brain has accurately generated a model inferring the causal structure of the external world (Friston, 2005, 2010). The notion we put forward gives specific constraints to the 'predictive coding' general framework.

A unique aspect of the MTM hypothesis is its attitude towards the conscious and unconscious perception dichotomy. The MTM model considers conscious perception as an ongoing process rather than a singular event in the brain. Accordingly, a stimulus is subjectively perceived as long as it is updated through a set of neuronal operations. If the processing chain is broken, then the information is not updated, and it echoes in the last processor until it fades. This, in turn, could influence behavior, although the observer would not be able to report it. Recent studies demonstrated that elaborate computations and manipulations could be done on stimuli even when they are rendered "Unseen". In these studies, subjects had successfully categorized unseen pictures, words and faces(S Dehaene et al., 2001; Qiao & Liu, 2009), managed to associate quantity to a number symbol (Naccache & Dehaene, 2001; Opstal, Gevers, Osman, & Verguts, 2010), a valence to a word (Kouider & Dehaene, 2007; Yeh, He, & Cavanagh, 2012). It seems that even manipulations that demand long maintenance of perceived information like solving an arithmetic series sum (Sklar et al., 2012) could be done on unconsciously perceived information (Dehaene, Charles, King, & Marti, 2014). These unexpected findings have dramatically extended the boundaries of unconscious processing which were originally thought to be restrained to simple operations. The model puts forward the possibility that the failure to report a stimulus means neither a complete absence of processing, nor a complete absence of subjective experience. If stimulus updating is interrupted, the existent representation of the stimulus or task-related sequence of processes could still affect behavior in the absence of a clear subjective experience at the time of the report. Since any part of the sequence of processes can fail, the remaining processes are not restrained to a specific level of complexity. In fact, this complexity might even strongly vary from trial to trial.

Traditionally, perception is viewed as a linear process in which unconscious perception precedes conscious perception. Accordingly, stimuli that were processed but did not gain conscious access could be considered as 'unconsciously perceived'. The MTM model does not exclude the possibility of unconscious perception. However, it predicts that multiple forms of 'degraded' perception can exist, resulting in distorted subjective experiences and variable effects on the observer's behavior. In accordance with this prediction, a recent study showed that one feature of a visual stimulus (e.g. color) can be consciously perceived while another aspect of the very same object can remain invisible (e.g. shape) (Elliott, Baird, & Giesbrecht, 2016). This prediction also fully agrees with the partial awareness hypothesis which suggests that stimulus features of various complexities can be consciously accessed independently (Kouider, de Gardelle, Sackur, & Dupoux, 2010).

## 4. Discussion
### Theories of consciousness and their relation to the current proposal

There are several prominent brain models of conscious perception which differ essentially in the activity they associate with conscious perception. Some propose a key role of early recurrent activity (<200 ms) between visual areas which enables the exchange of information between areas and group the perceptual information (Koch, Massimini, Boly, & Tononi, 2016; Lamme, 2006). By contrast, other models such as the global neuronal workspace attribute conscious perception to late activity (~ the P3b time window) that involves prefrontal cortices (Dehaene & Changeux, 2011; Stanislas Dehaene, Changeux, Naccache, Sackur, & Sergent, 2006). The model asserts that this manifests the settling of brain activity into a temporary metastable state of global activity during which content-specific information is shared between distant areas. High-order theory of consciousness (Lau & Rosenthal, 2011) also propose a crucial role of frontal areas in consciousness but it relates this late activity as a generic mental representation of the observer as being in a particular mental state (e.g. "I see a blue car"). Finally, the Integrated Information Theory (IIT) argues that the distributed activity (including parietal and frontal areas) represents an integration of stimuli's various features (Tononi, Boly, Massimini, & Koch, 2016). In this model, it is the causal properties of the brain, rather than a specific structure, that determines its level of consciousness.

All the above models relate the conscious representation of a stimulus as a discrete event and do not consider its ongoing integration in a changing context. The suggested model rejects the notion that there is a specific moment in time in which conscious and unconscious fully dissociate. It considers subjective perception as a continuous process that is affected by three main factors, namely: stimulus saliency, observer's goals and context. To maintain this ongoing process different brain regions are recruited (as anticipated by the GNW, IIT and high order models). In accordance with the GNW model, the current hypothesis puts forward the notion that the perceived stimuli are updated to a singular experience and therefore the underlying activity is not generic but is specific to the stimulus and the general context it is updated to. In accordance to the IIT model, information integration is a crucial part of the proposed model. However, our model highlights the qualitative aspects of this integration rather than its quantitative aspects. The kind of information provided by a neuronal population, rather than how much, should contribute to subjective experience. The idea that subjective contents evolve over time has been proposed to explain the apparent discrepancy in the timing of NCC (Bachmann, 2000). Albeit related, the hypothesis we are considering here is different: Here, the updating process is what makes and keeps the information conscious, whether the subjective experience is in an immature or highly detailed stage.

### Methodological implications

The comparison between seen and unseen stimuli in past studies lead to contradicting results. Different patterns of brain activity have been reported as NCCs (e.g. VAN, P3b, etc.). Researchers have debated this question for years, trying to avoid misestimating NCCs (see e.g. Tsuchiya et al., 2015). The current model suggests that the question of which brain process has the unique property of supporting subjective perception is ill-posed and thus has no answer. Instead, the model highlights the importance of moment to moment encoding, each brain area newly recruited modifying subjective experience. Accordingly, a stimulus is updated to an existing scene (context), depending on its saliency and its importance in accomplishing one or more of the observer's goals.

This theoretical shift obligates a respective methodological shift. The conscious/unconscious contrast could no longer be considered as a gold standard in extracting the neuronal activity that gives rise to subjective experience. One could no longer expect a clear-cut comparison that would extract a neural substrate or pinpoint a specific moment in time where conscious and unconscious perception diverge. Patterns of activity isolated by contrasting seen and unseen trials are expected to vary between studies (at least to a certain extent) because the processing stages related to the performed task, and the way they are disrupted in unseen trials, are different. Future studies should instead focus on the content of subjective experience and how it relates to the evolution of stimulus representation in the brain. Recent advances in signal processing and machine learning techniques make this goal achievable. The study by Salti et al. illustrates how these techniques allows tracking the processing of a stimulus in order to see how it is affected by experimental conditions.

## 5. Conclusion

We put forward a parsimonious model to account for subjective perception. Based on empirical evidence, we propose that subjective experience is built through a dynamic updating process. Accordingly, subjective perception does not correspond to a single neural event but instead is linked to a complex sequence of processes that vary with time, context, and subjects' goal. The suggested model narrows the gap between subjective experience and neural processing. Previous models assumed that some brain processes had a unique feature enabling subjective experience. The present model does not make such an assumption: it considers the possibility that successive processing steps directly contribute to subjective experience. This hypothesis bridges previously divergent interpretations of NCCs and provides new perspectives on the neural processes subtending subjective experience.